\documentclass{ws-ijmpc}

\begin{document}

\markboth{Julian Sienkiewicz, Janusz A. Ho{\l}yst}
{Scaling of internode distances in weighted complex networks}

\catchline{}{}{}{}{}

\title{SCALING OF INTERNODE DISTANCES IN WEIGHTED COMPLEX NETWORKS}

\author{JULIAN SIENKIEWICZ}

\address{Faculty of Physics\\
Center of Execellence for Complex Systems Research\\
Warsaw University of Technology, Koszykowa 75\\
Warsaw, 00-662, Poland\\
julas@if.pw.edu.pl}

\author{JANUSZ A. HO{\L}YST}

\address{Faculty of Physics\\
Center of Execellence for Complex Systems Research\\
Warsaw University of Technology, Koszykowa 75\\
Warsaw, 00-662, Poland\\
jholyst@if.pw.edu.pl}

\maketitle

\begin{history}
\received{Day Month Year}
\revised{Day Month Year}
\end{history}

\begin{abstract}
We extend the previously observed scaling equation connecting the internode distances and nodes' degrees onto the case of weighted networks. We show that the scaling takes a similar form in the empirical data obtained from networks characterized by different relations between node's strength and its degree. In the case of explicit equation for $s(k)$ (e.g. linear or scale-free), the new coefficients of scaling equation can be easily obtained. We support our analysis with numerical simulations for Erd\H{o}s-R\'{e}nyi random graphs with different weight distributions. 

\keywords{weighted networks; scaling; empirical data}
\end{abstract}

\ccode{PACS Nos.: 89.75.Hc, 89.75.Da}

\section{Introduction}

The are different ways to represent the structure of real-world networks using the notions of graph theory. In particular, depending on the fact if we are interested only in the question of existence of connections between nodes or we would also like to consider the "intensity" of those links, the described network can be presented as {\it unweighted} or {\it weighted}. In most cases the initial network is of weighted character but in simple analytical methods the weights are usually removed.

The modeling of weighted networks has begun almost parallel with the unweighted case \cite{yook_w}, however it is the higher availability of the empirical data that made possible to perform extensive and well set research on weighted transportation networks, networks of scientific collaboration \cite{scaling4} or in general the networks of social contacts \cite{kumpula_w}. In the same time the works on connections between topology and weight's dynamics \cite{wu_w,ramasco_w}, transport on weighted networks \cite{barrat_w,wang_w} and optimal paths \cite{chen_w} were being developed. Only recently the weighted networks approach has been used to create a unified statistics combing Bose-Einstein and Fermi-Dirac distributions \cite{garlaschelli_w}.

The considerations included in this paper are an attempt for extending the scaling relations presented in \cite{holyst_pre1,holyst_physa1} onto the case of weighted networks. In the cited works it is shown that the mean distance (or internode distance) between nodes with degrees $k_i$ and $k_j$ is given by the following equation
\begin{equation}
\langle l_{ij} \rangle = a - b \log k_i k_j,
\label{eq:lij}
\end{equation}
where a simple model of unweighted network with a negligible clustering coefficient leads to
\begin{equation}\label{eq:aib}
a=1+\frac{\log (N \langle k \rangle)}{\log
\kappa}\;\;\;\;\mbox{and}\;\;\;\; b=\frac{1}{\log \kappa}.
\end{equation}
with being $N$ the size of the network, $\langle k \rangle$ its average degree and $\kappa$
the average branching degree. The scaling relation (\ref{eq:lij}) has been spotted in many different networks ranging from public transport networks to biological ones\cite{holyst_pre1,holyst_physa1} although the observed values of the coefficients $a$ and $b$ are different from the results given by Eq. (\ref{eq:aib}).

\section{Empirical data}\label{edata}
In the case of weighted network instead of node's degree $k_i$ one usually considers node's strength $s_i$ i.e.
\begin{equation}
s_i = \sum_{j=1}^{j=k_i}w_{ij}
\label{eq:s}
\end{equation}
where $w_{ij}$ is the weight of the link between nodes $i$ and $j$. In order to find the relation between the internode distances $\langle l_{ij} \rangle$ and the product of nodes' strengths we examined the following systems: the network of flights between world-wide airports\footnote{The data contained information about $N=3073$ airports with average degree value $\langle k \rangle=10.3$ and they were obtained from the web-page of {\it OAG} http://www.oag.com corresponding to the status on 14th March 2006.}, public transport network in Warsaw\footnote{The network is described in \cite{sienkiewicz_pre1} with $N=1530$ stops and average degree $N=1530$ $\langle k \rangle=2.8$; data was downloaded form Warsaw Transportation Authorities web-page in April 2004.} and two networks made from connections between the web-pages of web portals\footnote{The data from two large Polish portals \cite{ania_pre} were obtained thanks to collaboration with Gemius company. For reasons of information confidentiality they will be further referred to as {\it portal A} (number of nodes $N=195$, average degree $\langle k \rangle=70$) and {\it portal B} ($N=512$, $\langle k \rangle=102$)}. In the case of the airport network the weight $w_{ij}$ stands for the number of different carriers making the connections between the airports $i$ and $j$ while the weights in Warsaw public transport network are proportional to the total number of buses and trams traveling between stops $i$ and $j$. In the web portals data the weight was defined as the total number of users who, being first at page $i$ surfed to page $j$. For simplicity reasons in all datasets the weights were symmetrized by calculating the average of the original values i.e. $w=(w_{ij}+w_{ji})/2$. Due to high reciprocity between $w_{ij}$ and $w_{ji}$ this does not affect further analysis.

\begin{figure*}[!ht]
\centerline{\epsfig{file=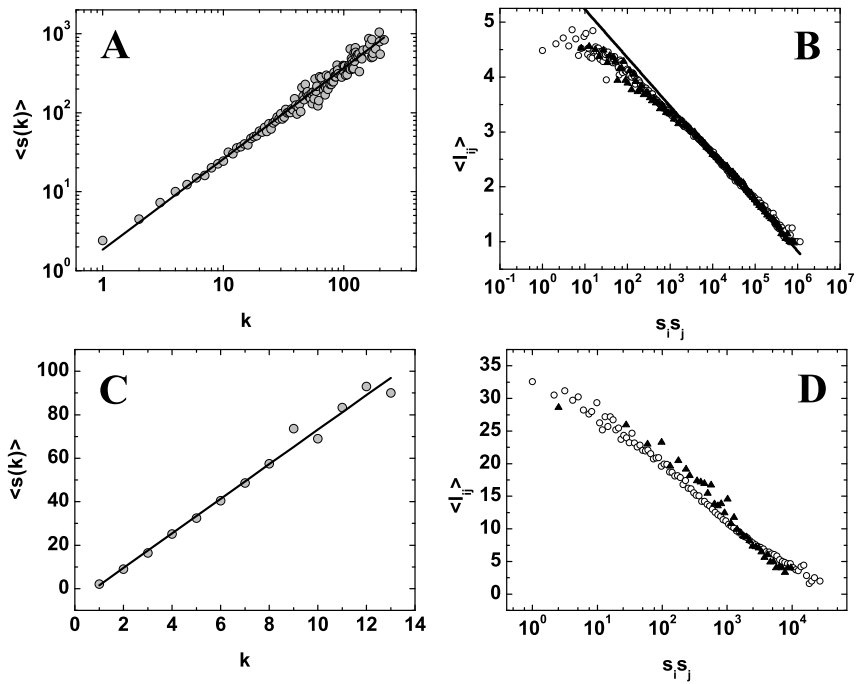,width=0.9\textwidth}}
\caption{The average node's strength versus node's degree $\langle s(k) \rangle$ for world-wide airport network (A) and public transport network in Warsaw (C). The solid line is a fit to the empirical data $\langle s(k) \rangle \sim k^{1.15}$ (A) and $\langle s(k) \rangle=8k-6$ (C). The dependence of the internode distances $\langle l_{ij} \rangle$ on the product of nodes' strengths $s_i s_j$ (circles) and on the product of nodes' degrees $k_i k_j$ rescaled by the relation $\langle s(k) \rangle$ (triangles) for the airport connections (B) and public transport in Warsaw (D). The solid line in (B) is defined by the rescaled coefficients $\widetilde{a}$ and $\widetilde{b}$ obtained from Eqs (\ref{eq:aibwsk}).}
\label{fig:wagi_transp}
\end{figure*}

\begin{figure*}[!ht]
\centerline{\epsfig{file=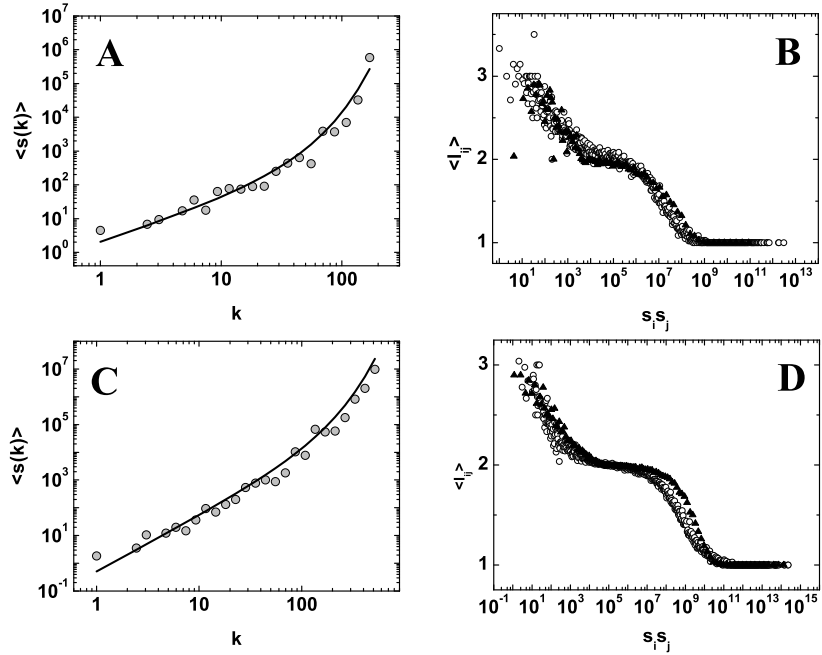,width=0.9\textwidth}}
\caption{
The average node's strength versus node's degree $\langle s(k) \rangle$ for Portal A (A) and portal B (C). The solid line is a fit to the empirical data $\langle s(k) \rangle = 2k^{1.2}\exp(k/30)$ (A) and $\langle s(k) \rangle = 0.5 k^2 \exp(k/100)$ (C). The dependence of the internode distances $\langle l_{ij} \rangle$ on the product of nodes' strengths $s_i s_j$ (circles) and on the product of nodes' degrees $k_i k_j$ rescaled by the relation $\langle s(k) \rangle$ (triangles) for the Portal A (B) and Portal B (D).}
\label{fig:wagi_int}
\end{figure*}

The analysis of the results obtained from the above described datasets is presented in Figures \ref{fig:wagi_transp} and \ref{fig:wagi_int}. For each network we plot the dependence of average node's strength and its degree $\langle s(k) \rangle$ (Figures \ref{fig:wagi_transp}A, \ref{fig:wagi_transp}C, \ref{fig:wagi_int}A and \ref{fig:wagi_int}C) as well as the internode distances $\langle l_{ij} \rangle$ separately as a function of the product of degrees $k_i k_j$ and as a function of the product of strengths $s_i s_j$ (Figures \ref{fig:wagi_transp}B, \ref{fig:wagi_transp}D, \ref{fig:wagi_int}B and \ref{fig:wagi_int}D). 

The plots presenting $\langle l_{ij} \rangle$ as a function of $k_i k_j$ were rescaled with respect to node degree values, substituting it with an approximated relation between the node strength and its degree, marked as solid curve in $\langle s(k) \rangle$ plots. The $\langle s(k) \rangle$ equation can take different forms: starting from lack of correlations in case of public transport network in Warsaw ($\langle s(k) \sim k$), through a scale-free relation in airport network ($\langle s(k) \rangle \sim k^{\alpha}$), ending at a complicated dependence seen in web portals data ($\langle s(k) \rangle \sim k^{\alpha}\exp(\beta k)$). Nevertheless in each example the coincidence between the rescaled data and the empirically observed relation $\langle l_{ij} (s_i s_j)\rangle$is clearly visible. The oscillations observed on the logarithmic trend seen in Figures \ref{fig:wagi_int}B and \ref{fig:wagi_int}D are an effect of the high average degree present in both portals data. The origin of this phenomenon is extensively described in \cite{sienkiewicz_pre2}.

\section{Scaling in the absence of $s(k)$ correlations}
\begin{figure*}[!ht]
\centerline{\epsfig{file=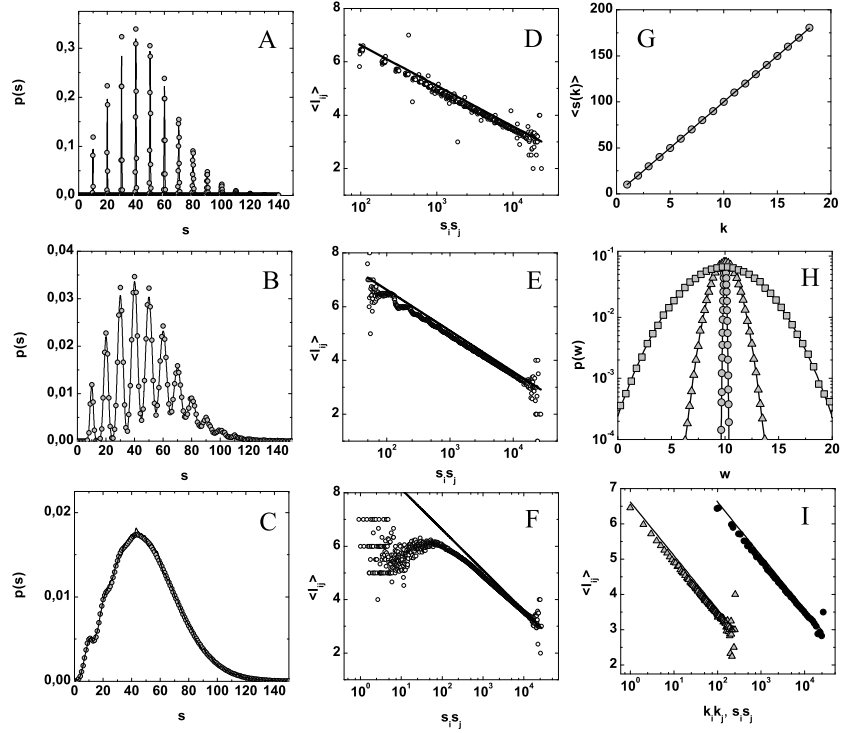,width=0.9\textwidth}}
\caption{Numerical simulations of the ER network with $N=1000$ and $\langle k \rangle=5$ and the weights coming from Gauss distribution (\ref{eq:wgauss}) with $\langle w \rangle=10$ and different values of standard deviation $\sigma_w$. (A--C) Strength distributions for $\sigma_w=0.1$ (A), $\sigma_w=1$ (B) and $\sigma_w=3$ (C); solid lines correspond to numerically summed Eq. (\ref{eq:psssuma}). (D--F) Internode distances versus the product of strengths for $\sigma_w=0.1$ (D), $\sigma_w=1$ (E) and $\sigma_w=3$ (F); solid lines correspond to  Eq. (\ref{eq:lss}). (G) The node's strength versus its degree; solid line corresponds to equation $s(k)=10k$. (H) Weight distributions for $\sigma_w=0.1$ (circles), $\sigma_w=1$ (triangles) and $\sigma_w=3$ (squares); solid lines come from Eq. (\ref{eq:wgauss}). (I) A comparison  between relations $\langle l_{ij} \rangle (s_i s_j)$ (circles) and $\langle l_{ij} \rangle (k_i k_j)$ (triangles) for $\sigma_w=0.1$; solid lines come from Eqs (\ref{eq:lij}) and (\ref{eq:lss}).}
\label{fig:wagi_gen_gauss}
\end{figure*}

The scaling relation observed in the absence of strength-degree correlations can be easily explained using the approach previously applied for internode distances in \cite{holyst_pre1}. In fact, considering a branching process one obtains the following equation connecting the degrees of nodes $i$ and $j$, the distance $x$ between them, the size of the network $N$, its average degree $\langle k \rangle$ and the average branching degree $\kappa$:
\begin{equation}
k_i \kappa^{x-1} = \frac{N \langle k \rangle}{k_j}.
\label{eq:NM}
\end{equation}
Multiplying the nominator and denominator at both sizes of the equation by the average value of the weight in the network $\langle w \rangle$ and assuming that in the absence of the $s(k)$ correlations the following relation is satisfied
\begin{equation}
s_i = \langle w \rangle k_i,
\label{eq:sk}
\end{equation}
one gets
\begin{equation}
s_i \kappa^{x-1} = \frac{N \langle k \rangle \langle w \rangle^2}{s_j}.
\label{eq:NMs}
\end{equation}
This equation immediately results in the form of the internode distances scaling relation as a function of nodes' strengths 
\begin{equation}
\langle l_{ij} \rangle = \widetilde{a} - \widetilde{b} \log s_i s_j,
\label{eq:lss}
\end{equation}
where coefficients $\widetilde{a}$ and $\widetilde{b}$ are given by
\begin{equation}\label{eq:aibw}
\widetilde{a}=1+\frac{\log (N \langle k \rangle \langle w \rangle^2)}{\log
\kappa}\;\;\;\;\mbox{oraz}\;\;\;\; \widetilde{b}=\frac{1}{\log \kappa}.
\end{equation}
The above considerations lead to the following conclusions: (a) the slope $\widetilde{b}$ of the scaling function is the same as in the unweighted case, (b) the relation between coefficients $a$ and $\widetilde{a}$ takes the form
\begin{equation}
\widetilde{a} = a + \frac{\log \langle w \rangle^2}{\log \kappa}.
\label{eq:aaw}
\end{equation}
In order to compare the consistency of the Eqs (\ref{eq:lss}) and (\ref{eq:aibw}) with the numerical simulations we performed the trails (see Figure \ref{fig:wagi_gen_gauss}) for the Erd\H{o}s-R\'{e}nyi random network (ER network) of $N=1000$ nodes and $\langle k \rangle=5$. The weights were randomly drawn from the Gauss distribution
\begin{equation}
p_w(w)=\frac{1}{\sqrt{2\pi}\sigma_w}\mathrm{e}^{-\frac{(w-\langle w \rangle)^2}{2 \sigma^2_w}},
\label{eq:wgauss}
\end{equation}
characterized by the mean value $\langle w \rangle=10$ and three different values of the standard deviation: $\sigma_w=0.1$ (Figures \ref{fig:wagi_gen_gauss}A and \ref{fig:wagi_gen_gauss}D), $\sigma_w=1$ (Figures \ref{fig:wagi_gen_gauss}B and \ref{fig:wagi_gen_gauss}E) and $\sigma_w=3$ (Figures \ref{fig:wagi_gen_gauss}C and \ref{fig:wagi_gen_gauss}F). A comparison of those weight distributions is presented in Figure \ref{fig:wagi_gen_gauss}H. As one can see the width of the distribution plays the key role with respect to the shape of the strength distribution $P(s)$; in case when $\sigma_w=0$ one gets of course the Poisson distribution
\begin{equation}
P(s)=\mathrm{e}^{-\langle k \rangle} \frac{\langle k \rangle^{\frac{s}{\langle w \rangle}}}{\left( \frac{s}{\langle w \rangle} \right)!},
\end{equation}
for $s=\langle w \rangle$,$2 \langle w \rangle$,$3 \langle w \rangle,...$. In general, the strength distribution takes the form
\begin{equation}
P(s)=\sum_{k=1}^{\infty}p(k) \rm{Prob} \left( \sum_{i=1}^{k}w_i =s\right),
\end{equation}
assuming that weights $w_i$ are i.i.d. random variables drawn from the distribution $p_w(w)$. 
Taking into account the considered distributions $p(k)$ and $p_w(w)$ one arrives at the relation
\begin{equation}
P(s) = \sum_{k=1}^{\infty}\frac{1}{\sqrt{2\pi k}\sigma_w}\mathrm{e}^{-\frac{(s-\langle w \rangle k)^2}{2 k \sigma^2_w}}\mathrm{e}^{-\langle k \rangle} \frac{\langle k \rangle^k}{k!}. 
\label{eq:psssuma}
\end{equation}
The strength distributions taken from the numerical simulations for three different values of $\sigma_w$ as well as the values $P(s)$ numerically summed using  Eq. (\ref{eq:psssuma}) are shown in Figures \ref{fig:wagi_gen_gauss}A, \ref{fig:wagi_gen_gauss}B and \ref{fig:wagi_gen_gauss}C. The characteristic oscillations seen in $P(s)$ distribution around the multiply of the average weight ($\langle w \rangle$,$2 \langle w \rangle$,$3 \langle w \rangle$,...) vanish along with the increasing value of $\sigma_w$ which in result gives the distribution a smooth shape. The form of the strength distribution changes the relation between the internode distances and the product of strengths (see Figures \ref{fig:wagi_gen_gauss}D--F). The relation bewteen the average node's strength and node's degree follows Eq. (\ref{eq:sk}) and is shown in Figure \ref{fig:wagi_gen_gauss}G.

For $\sigma_w=0.1$ (Figure \ref{fig:wagi_gen_gauss}D) the points obtained from the numerical simulations overlap those predicted by Eq. (\ref{eq:lss}) with the coefficients $\widetilde{a}$ and $\widetilde{b}$ defined by the parameters of the ER network and the average weight $\langle w \rangle$. A comparison  between relations $\langle l_{ij} \rangle (s_i s_j)$ and $\langle l_{ij} \rangle (k_i k_j)$ is presented in Figure \ref{fig:wagi_gen_gauss}I showing that indeed the slopes in both cases are identical as predicted by Eqs (\ref{eq:aib}) and (\ref{eq:aibw}). The increase in the $\sigma_w$ value leads to emergence of the oscillations on the $log(s_i s_j)$ trend line (Figure \ref{fig:wagi_gen_gauss}E) and eventually for large values of $\sigma_w$ the scaling relations breaks down. This effect is illustrated in Figure \ref{fig:wagi_gen_gauss}F.

\section{Scaling in the presence of $s(k)$ correlation}
Section \ref{edata} contains examples of the scaling of internode distances on the product of nodes' strengths in the presence of non-linear $s(k)$ correlations. Previously proposed branching-tree model has no application in this case, as one should sum up all the weights on the tree's surface, which is impossible for any general form of $s(k)$. Certainly, if $s(k)$ has an explicit form, there is a possibility of reversing the relation, putting it into Eq. (\ref{eq:NM}) and obtaining the coefficients $\widetilde{a}$ and $\widetilde{b}$ defining the $\langle l_{ij} \rangle$ relation as a function of $s_i s_j$. For instance, if $s(k) = Ak^{\alpha}$, reversing gives $k = (s/A)^{1/\alpha}$ and as a result the rescaled coefficients $\widetilde{a}$ and $\widetilde{b}$ will take the form
\begin{equation}\label{eq:aibwsk}
\widetilde{b}=\frac{b}{\alpha}\;\;\;\;\mbox{and}\;\;\;\; \widetilde{a}=a + 2 \log A \widetilde{b}.
\end{equation}
Such an example is presented in Figure \ref{fig:wagi_transp}B (solid line) where, after getting coefficients $a$ and $b$ from fitting to relation (\ref{eq:lij}) they were put into Eqs \ref{eq:aibwsk}) thus obtaining the values $\widetilde{a}$ and $\widetilde{b}$.

\begin{figure*}[!ht]
\centerline{\epsfig{file=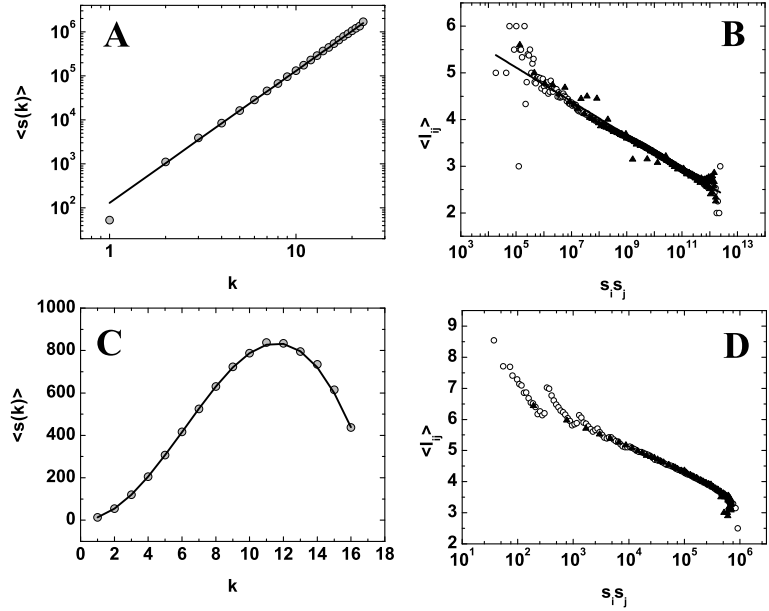,width=0.8\textwidth}}
\caption{Plots {\it A} and {\it C}: the average node's strength versus its degree $k$. Plots {\it B} and {\it D}: the internode distance versus the product of nodes' degrees $k_i k_j$ rescaled by the relation between node's strength and its degree (triangles);  the internode distance versus the product of nodes' strengths $s_i s_j$ (circles). Plots A and B were obtained for ER network with $N=1000$, $\langle k \rangle=5$ where weights fulfill the equation $w_{ij} \sim (k_i k_j)^{2}$, while in the case of plots C and D a corresponding relation was $w_{ij} \sim \sin(k_i) \sin(k_j)$.}
\label{fig:wagi_gen}
\end{figure*}

For more systematical study we performed numerical simulations for ER network with $N = 1000$ nodes and $\langle k \rangle = 5$, with specific $s(k)$ relations. To generate the correlations between node's strength and its degree we used the observations from work \cite{scaling4}: if in the given network there are no degree-degree correlations (the asortativity coefficient $r$ is equal to 0) and the edge between nodes $i$ and $j$ has weight $\langle w_{ij} \rangle$, proportional to the expression $(k_i k_j)^{\alpha}$ then the strength of node $i$ is 
\begin{equation}\label{eq:aibww}
s_i \sim k_i \langle w_{ij} \rangle \sim k_i^{1+\alpha}.
\end{equation}
Using the above described procedure, we imposed weights onto ER network according to relations $s(k)=130 k^3$ (Figure \ref{fig:wagi_gen}A) and $s(k)=80 k \sin(\pi k /18)$ (Figure \ref{fig:wagi_gen}C). As one can observe in Figures \ref{fig:wagi_gen}B and \ref{fig:wagi_gen}D the expected overlap of relation $\langle l_{ij} \rangle$ as a function of $s_i s_j$ (circles) with the rescaled data of $\langle l_{ij} \rangle (k_i k_j)$ (triangles) was obtained. Moreover, in Figure \ref{fig:wagi_gen}B there can be seen a good fitting of data with linear function determined by the coefficients taken from the Eqs (\ref{eq:aibwsk}).

\section{Conclusions}
The relation between internode distances and the product of the nodes' strengths is an extension of the previously examined dependence linking $\langle l_{ij} \rangle$ and $k_i k_j$. As expected the scaling takes a similar form while the coefficients $a$ and $b$ are altered by the relation between average node's strength and its degree. In the case of explicit equation for $s(k)$ (e.g. linear or scale-free), the new coefficients $\widetilde{a}$ and $\widetilde{b}$ can be easily obtained. A breakdown in scaling relation while increasing the width of the weight distribution has also been observed.

\section*{Acknowledgments}
This work was supported by Polish Ministry of Science Grant 1750/B/H03/2008/35.

\end{document}